\documentclass[pra,twocolumn,superscriptaddress,amsmath,amssymb,floatfix]{revtex4-1}
\usepackage{graphicx}
\usepackage{url}

\begin{document}

\title{Optical conductivity of multifold fermions: the case of RhSi}

\author{L. Z. Maulana}
\affiliation{1. Physikalisches Institut, Universit\"at Stuttgart,
70569 Stuttgart, Germany}
\author{K. Manna}
\affiliation{Max-Planck-Institut f\"ur Chemische Physik fester
Stoffe, 01187 Dresden, Germany}
\author{E. Uykur}
\affiliation{1. Physikalisches Institut, Universit\"at Stuttgart,
70569 Stuttgart, Germany}
\author{C. Felser}
\affiliation{Max-Planck-Institut f\"ur Chemische Physik fester
Stoffe, 01187 Dresden, Germany}
\author{M. Dressel}
\affiliation{1. Physikalisches Institut, Universit\"at Stuttgart,
70569 Stuttgart, Germany}
\author{A. V. Pronin}
\email{artem.pronin@pi1.physik.uni-stuttgart.de} \affiliation{1.
Physikalisches Institut, Universit\"at Stuttgart, 70569 Stuttgart,
Germany}

\date{April 9, 2020}

\begin{abstract}

We measured the reflectivity of the multifold semimetal RhSi in a
frequency range from 80 to 20\,000~cm$^{-1}$ (10 meV -- 2.5 eV) at
temperatures down to 10~K. The optical conductivity, calculated from
the reflectivity, is dominated by the free-carrier (Drude)
contribution below 1000~cm$^{-1}$ (120 meV) and by interband
transitions at higher frequencies. The temperature-induced changes
in the spectra are generally weak: only the Drude bands narrow upon
cooling, with an unscreened plasma frequency that is constant with
temperature at approximately 1.4~eV, in agreement with a weak
temperature dependence of the free-carrier concentration determined
by Hall measurements. The interband portion of conductivity exhibits
two linear-in-frequency regions below 5000~cm$^{-1}$ ($\sim$~600
meV), a broad flat maximum at around 6000~cm$^{-1}$ (750 meV), and a
further increase starting around 10\,000~cm$^{-1}$ ($\sim$~1.2 eV).
We assign the linear behavior of the interband conductivity to
transitions between the linear bands near the band crossing points.
Our findings are in accord with the predictions for the low-energy
conductivity behavior in multifold semimetals and with earlier
computations based on band structure calculations for RhSi.

\end{abstract}

\maketitle

\section{Introduction}

Multifold fermions are quasiparticles described by higher-spin
generalizations of the Weyl equation. They can be realized in the
multifold semimetals: materials, which possess characteristic
electronic band crossings with degeneracies higher than
two~\cite{Manes2012, Bradlyn2016}. A number of such semimetals were
recently predicted and experimentally confirmed among the materials
from the space group 198 (SG198), whose symmetry is
noncentrosymmetric and has no mirror planes, leading to a
realization of ``topological chiral crystals''~\cite{Chang2018,
Chang2017, Tang2017, Sanchez2019, Rao2019, Schroter2019,
Takane2019}. In such semimetals, the quantized circular
photogalvanic effect (QCPGE) was forecasted in
2017~\cite{deJuan2017}. In this non-linear optical phenomenon,
helical (i.e., circularly polarized) photons excite the chiral band
carriers in such a way that the resultant photocurrent is quantized
in units of material-independent fundamental constants. Recently,
the observation of QCPGE was reported in RhSi~\cite{Rees2019}, a
member of SG198 and an established multifold
semimetal~\cite{Chang2017, Tang2017, Sanchez2019}.

These remarkable theoretical results and experimental observations
demand further optical characterizations of RhSi, in particular
since the knowledge of frequency-dependent conductivity,
$\sigma(\omega)=\sigma_{1}(\omega)+i\sigma_{2}(\omega)$, is
essential for proper interpretation of QCPGE experiments. Because
there is no characteristic energy scale, the quasiparticles near the
linear-band crossings in three dimensions are expected to be
characterized by an interband $\sigma_{1}(\omega)$ that is
proportional to the probing light frequency~\cite{Hosur2012,
Bacsi2013, Ashby2014}. Indeed, such linear $\sigma_{1}(\omega)$ has
been observed in many Dirac and Weyl semimetals~\cite{Chen2015,
Xu2016, Neubauer2016, Kimura2017}. The optical conductivity of the
multifold semimetals has also been predicted to demonstrate a linear
$\sigma_{1}(\omega)$ \cite{Grushin2019}. Most recently, optical
conductivity was specifically calculated for RhSi from its band
structure~\cite{Li2019}. In this paper, we experimentally examine
these theoretical results.

\section{Experiment}

Single crystals of RhSi were grown in the same way as described in
Ref.~\cite{Rees2019}. The vertical Bridgman crystal-growth technique
was utilized to grow the crystals from the melt using a slightly
off-stoichiometric composition (excess Si). First, a polycrystalline
ingot was prepared using the arc-melt technique by mixing the
stoichiometric amount of constituent Rh and Si elements of 99.99\%
purity. Then the crushed powder was filled in a custom-designed
sharp-edged alumina tube and finally sealed inside a tantalum tube
with argon atmosphere. A critical composition with slightly excess
Si was maintained to ensure a flux growth inside the Bridgman
ampule. The whole assembly was heated to 1550~$^{\circ}$C with a
rate of 200~$^{\circ}$C per hour and halted for 12 hours to ensure
good mixing of the liquid. Then the crucible was slowly pulled to
$\sim$~1100~$^{\circ}$C with a rate of 0.8 mm/h and finally quenched
to room temperature. The temperature profile was controlled by
attaching a thermocouple at the bottom of the tantalum ampule
containing the sample. Single crystals with average linear
dimensions of a few millimeters were obtained. The crystals were
first analyzed with a white beam backscattering Laue x-ray
diffractometer at room temperature. The obtained single and sharp
Laue spot could be indexed by a single pattern, revealing excellent
quality of the grown crystals without any twinning or domains. The
structural parameters were determined using a Rigaku AFC7
four-circle diffractometer with a Saturn 724+ CCD detector applying
graphite-monochromatized Mo-$K\alpha$ radiation. The crystal
structure was refined to be cubic $P2_{1}3$ (SG198) with the lattice
parameter $a$=4.6858(9)~{\AA}.

\begin{figure}[t]
\centering
\includegraphics[width=0.9\columnwidth,clip]{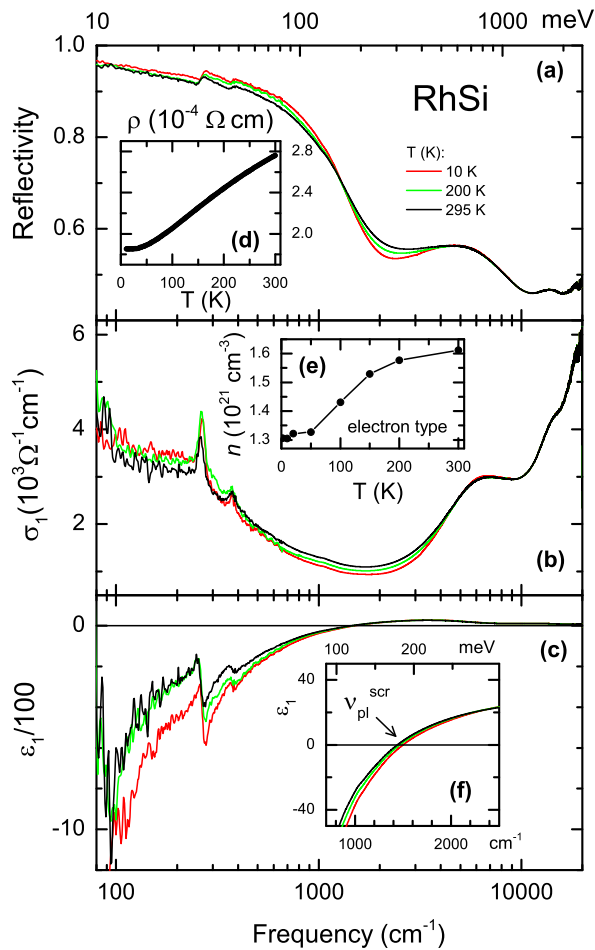}
\caption{Optical reflectivity (a) and the real parts of the optical
conductivity (b) and the dielectric permittivity (c) of RhSi at
selected $T=10$, $200$, and $295$~K. Note the logarithmic $x$ scale.
The insets show: the dc resistivity vs. $T$ (d), the Hall electron
concentration vs. $T$ (e), and a zoom of the permittivity spectra
near the zero crossings (f).} \label{overview}
\end{figure}

Temperature-dependent transport measurements (longitudinal dc
resistivity and Hall) were performed in a custom-made setup at
temperatures down to 2 K. The results of these measurements are
displayed in Figs.~\ref{overview}(d) and \ref{overview}(e). A
metallic behavior with a residual resistivity of $1.86 \times
10^{-4}$~$\Omega$cm was observed. The Hall measurements evidenced
electron conduction with a moderate increase of carrier
concentration upon increasing temperature.

\begin{figure}[t]
\centering
\includegraphics[width=8 cm,clip]{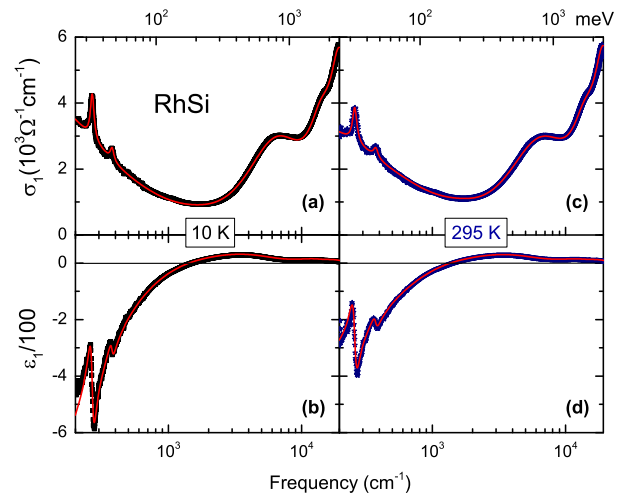}
\caption{Drude-Lorentz fits (lines) of the measured optical spectra
(symbols) of $\sigma_{1}$ and $\varepsilon_{1}$ of RhSi at 10 and
295 K, as indicated. Note the logarithmic $x$ scale.} \label{fit}
\end{figure}

Optical reflectivity, $R(\nu)$, was measured on a polished surface
of a RhSi single crystal (with roughly $1.5 \times 1.5$ mm$^{2}$ in
lateral dimensions) over a broad frequency range from $\nu =
\omega/(2 \pi)= 80$ to 20\,000 cm$^{-1}$ (10 meV -- 2.5 eV) at
several different temperatures ($T$ = 10, 25, 50, 75, 100, 150, 200,
250, 295 K). The spectra in the far infrared (below 700 cm$^{-1}$)
were collected with a Bruker IFS 113v Fourier-transform spectrometer
using \textit{in situ} gold coating of the sample surface for
reference measurements~\cite{gold}. At higher frequencies, a Bruker
Hyperion infrared microscope attached to a Bruker Vertex 80v
Fourier-transform infrared spectrometer was used. For these
measurements, freshly evaporated gold mirrors served as reference.
No sample anisotropy was detected in agrement with the cubic
crystallographic structure. Justification of using polished sample
surfaces is provided in the Supplemental Material below.

For Kramers-Kronig analysis, zero-frequency extrapolations were made
using the Hagen-Rubens relation in accordance with the
temperature-dependent longitudinal dc resistivity measurements. For
high-frequency extrapolations, we utilized the x-ray atomic
scattering functions~\cite{Tanner2015} followed by the free-electron
behavior, $R(\omega) \propto 1/\omega^{4}$, above 30 keV.

We found that the skin depth of the probing radiation exceeds 30 nm
for all temperatures and frequencies (in the far-infrared range, it
is above 200 nm). Hence, our optical measurements reflect the bulk
properties of RhSi.

\section{Results and discussion}

Figure~\ref{overview} displays the optical reflectivity $R(\nu)$ and
the real parts of the optical conductivity and dielectric constant,
$\varepsilon_{1}(\nu)=1-2\sigma_{2}(\nu)/\nu$, for the studied RhSi
sample in the full measurement-frequency range at three different
temperatures. The overall temperature evolution of the spectra is
minor. For frequencies higher than $\sim$ 8\,000~cm$^{-1}$ ($\sim 1$
eV), the optical properties are fully independent of temperature.
For the spectra analysis, we performed standard Drude-Lorentz
fits~\cite{Dressel2002}, where the Drude terms describe the
free-carrier response, while the Lorentzians mimic the interband
optical transitions and phonons. Examples of such fits are presented
in Fig.~\ref{fit}. Here we kept the zero-frequency limit of optical
conductivity equal to the measured dc-conductivity value at every
temperature. No other restrictions on the fit parameters were
imposed. For best possible model description, the experimental
spectra of $R(\nu)$, $\sigma_{1}(\nu)$ and $\varepsilon_{1}(\nu)$
were fitted simultaneously.

\subsection{Electronic response}

\begin{figure}[t]
\centering
\includegraphics[width=0.8\columnwidth,clip]{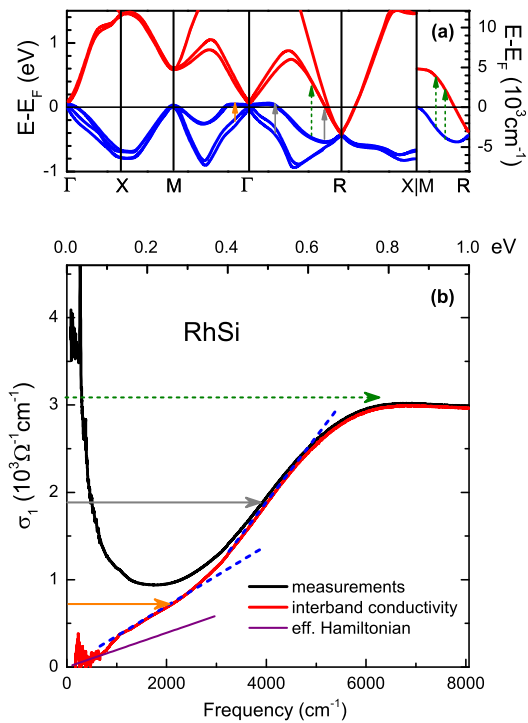}
\caption{Low-energy electronic structure of RhSi (a) and its optical
conductivity (b). The electronic structure is calculated within the
Topological Materials Database project~\cite{Bradlyn}. Note that the
band color code follows this reference and is not always related to
the band position relative to the Fermi level. Both, the as-measured
conductivity spectra and the spectra after subtraction of the
free-carrier response, $\sigma_{1}^{\rm{inter}}(\nu)$, are shown in
(b). The optical transitions responsible for the characteristic
features in $\sigma_{1}^{\rm{inter}}(\nu)$ are depicted as vertical
arrows in (a). The corresponding frequency scales are indicated by
the horizontal arrows of the same color in (b). The solid orange
arrows represent the transitions between the approximately linear
bands (including the almost flat bands) near the $\Gamma$ point and
the corresponding linear $\sigma_{1}^{\rm{inter}}(\nu)$. The solid
grey arrows indicate two different processes coinciding in
frequency: the onset of the downturn of the flat bands and the onset
of interband transitions in the vicinity of the $R$ point. These
processes lead to additional contributions to $\sigma_{1}(\nu)$. The
dashed green arrows correspond to the transitions between the almost
parallel bands along the $M$--$R$ line, leading to a maximum in
$\sigma_{1}(\nu)$. Transitions with approximately, but not exactly,
the same energy (e.g., along the $\Gamma$--$R$ line) are responsible
for smearing out this maximum and are indicated with the same arrow.
The dashed blue lines are a guide to the eye. The solid purple line
is an extrapolation of the effective-Hamiltonian computations from
Ref.~\cite{Grushin2019} to higher frequencies.} \label{bands}
\end{figure}

At low frequencies, RhSi demonstrates a typical (semi)metallic
response. The intraband contribution to the spectra can be best
fitted with two Drude components, which have different scattering
rates. Such multi-component Drude fits are often used to describe
the optical response of multiband systems, particularly different
semimetals~\cite{Schilling2017, Neubauer2018, Qiu2019}. In the case
of RhSi, the two-Drude approach can be justified by the presence of
a few bands crossing the the Fermi level~\cite{Chang2017, Tang2017,
Li2019, Bradlyn}; see Fig.~\ref{bands}(a): one set of bands is
around the $R$ point and the others are near the $\Gamma$ and $M$
points of the Brillouin zone. The first set provides the dominating
contribution to the free-carrier response and is also responsible
for the electron type of conduction. Still, the second Drude term is
necessary to describe the optical spectra accurately. Let us note
that exact interpretation of such a two-Drude approach is arguable.
The two components are not necessarily associated with two different
bands, but might instead be related to the scattering processes
within a band and between two different bands with a phonon
involved.

\begin{figure}[b]
\centering
\includegraphics[width=\columnwidth,clip]{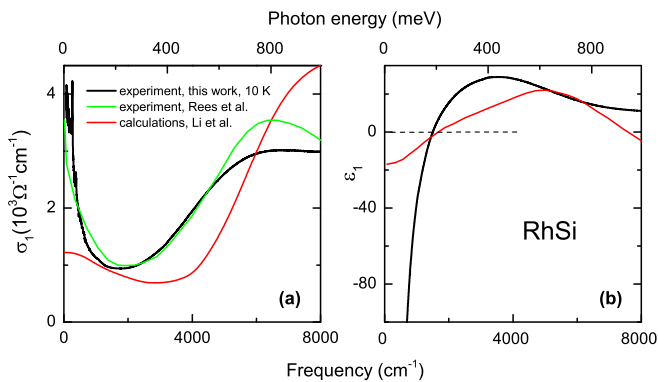}
\caption{Comparison of the experimental (this work and
Ref.~\cite{Rees2019}) and calculated (Ref.~\cite{Li2019}) optical
spectra of RhSi: optical conductivity (a) and dielectric function
(b).} \label{comparison}
\end{figure}

The scattering rates of the Drude terms are found to be 250
cm$^{-1}$ ($\sim$~30 meV) and 800 cm$^{-1}$ (100 meV) at 100~K.
These values correspond to the midranges for the spectra at all
measurement temperatures. At $T$ different from 100~K, the
scattering rates change only within $\pm 20$~\% of these values. The
mid-range relaxation times are, hence, 21 and 6.6 fs. These values
are somewhat larger than the value reported in another optical study
of RhSi~\cite{Rees2019} indicating the improved quality of RhSi
samples investigated in this work.

At around 1500~cm$^{-1}$ ($\sim 200$ meV), a characteristic plasma
edge is observed in $R(\nu)$. This edge correlates with the zero
crossing of $\varepsilon(\nu)$, which corresponds to the screened
plasma frequency, $\nu_{pl}^{scr} = \nu_{pl} /
\sqrt{\varepsilon_{\rm{inf}}}$. Here, $\varepsilon_{\rm{inf}} = 65
\pm 5$, is the cumulative contribution of the higher-frequency
optical transitions to $\varepsilon_{1}$ as obtained from the fits
and $\nu_{pl}$ is the unscreened plasma frequency. As best seen from
Fig.~\ref{overview}(f), the screened plasma frequency is
$\nu_{pl}^{scr} = (1470 \pm 30)$ cm$^{-1}$ [$\hbar\omega_{pl}^{scr}
= (182 \pm 4)$ meV] and is independent of temperature, the
corresponding unscreened plasma frequency being $\nu_{pl} = (11\,900
\pm 700)$ cm$^{-1}$ [$\hbar\omega_{pl} = (1470 \pm 90)$ meV]. This
value of plasma frequency coincides within the experimental
uncertainty with the value obtained from the Drude fits, $\nu_{pl} =
(11\,200 \pm 600)$ cm$^{-1}$ [$\hbar\omega_{pl} = (1390 \pm 80)$
meV], which includes contributions from both Drude terms and also
shows no $T$ dependence.

The absence of any detectable temperature dependence of
$\omega_{pl}$ is in qualitative agreement with the very modest
temperature-induced change of the carrier concentration: $n(T)$
increases by only 20~\% as $T$ goes from 2 to 300 K; see
Fig.~\ref{overview}(e).

The relatively large value of $\omega_{pl}$ (cf. the results for
other nodal semimetals~\cite{Chen2015, Xu2016, Neubauer2016,
Kimura2017, Hutt2018}) and the fairly high ($\sim10^{21}$cm$^{-3}$)
free-electron density of RhSi [see Fig.~\ref{overview}(d)] are
consistent with the results of band structure
calculations~\cite{Chang2017, Tang2017, Li2019, Bradlyn}, which show
that the Fermi level in RhSi is quite deep in the conduction band
for the electron momenta near the corners ($R$ points) of the
Brillouin zone [Fig.~\ref{bands}(a)]. This situation makes the
optical response of RhSi similar to the one observed in the Dirac
semimetal Au$_{2}$Pb~\cite{Kemmler2018}: free carriers dominate the
low-frequency ($\nu < 2000$ cm$^{-1}$, $\hbar\omega < 250$ meV)
regions of $\sigma_{1}(\nu)$ and $\varepsilon(\nu)$. Still, unlike
the situation in Au$_{2}$Pb, the optical transitions between the
linearly dispersing bands can be resolved in RhSi.

In Fig.~\ref{comparison} our optical findings at $T=10$~K are
plotted together with the results of band-structure-based
calculations from Ref.~\cite{Li2019} and with the previously
reported measurements of Ref.~\cite{Rees2019}. The experimental
curves follow each other quite well. The deviations between the
curves can be explained by different free-carrier contributions (cf.
the difference in the scattering times discussed above) and probably
by a somewhat more accurate Kramers-Kronig analysis utilized in the
present work: our reflectivity measurements are performed in a
broader frequency range as compared to the measurements from
Ref.~\cite{Rees2019}.

Despite some discrepancy between the calculations and both
experimental curves in Fig.~\ref{comparison}(a), the match can be
considered satisfactory. One has to keep in mind that calculations
of the optical conductivity from the electronic band structure are
rather challenging, particularly for semimetals: a survey of the
available literature reveals only a qualitative match between the
calculated optical conductivity and experimental results for a wide
range of nodal semimetals studied recently~\cite{Kimura2017,
Neubauer2018, Frenzel2017, Chaudhuri2017, Grassano2018}.
Nevertheless, both the low-energy features of the interband
experimental $\sigma(\nu)$ -- the initial (i.e., for the frequencies
just above the Drude rolloff) linear increase and the further
flattening -- are reproduced by theory.

In order to establish a better connection between the features
observed in the most interesting low-energy part of the experimental
conductivity and the interband optical transitions, we show our
$\sigma(\nu)$ together with the low-energy band structure of RhSi in
Fig.~\ref{bands}. Additionally, we plot the interband contribution
to the optical conductivity, $\sigma_{1}^{\rm{inter}}(\nu)$,
obtained by subtracting the Drude fits and the sharp phonon peaks
(discussed below) from the measured spectra. The interband optical
conductivity found this way is pretty much linear in frequency for
$\nu < 3000$~cm$^{-1}$. We found that this approximate linearity is
robust: varying the fit parameters within the uncertainty, set by
the experimental spectra, does not compromise it. (For details,
refer to the Supplemental Material in the end of the paper.)

At the lowest frequencies (below approximately 2500 cm$^{-1}$ or 0.3
eV), the interband conductivity is entirely due to transitions in
the vicinity of the $\Gamma$ point. No other interband optical
transitions are possible (either the direct gap between the bands is
too large, or the transitions are Pauli blocked). The bands near the
$\Gamma$ point are all roughly linear (two of them are basically
flat), thus a linear-in-frequency interband conductivity is
expected~\cite{Grushin2019}. Indeed, as already noticed,
$\sigma_{1}^{\rm{inter}}(\nu)$ is proportional to frequency in this
range (marked with the orange arrows). At somewhat higher $\nu$
($\sim$~$3000-4000$~cm$^{-1}$, $0.4-0.5$ eV), the flat bands start
to disperse downward, thus the linearity of
$\sigma_{1}^{\rm{inter}}(\nu)$ is not expected anymore. However, the
interband contributions in the vicinity of the $R$ points become
allowed at roughly the same energy [cf. the two grey arrows in panel
(a)]. These transitions provide a dominating contribution to
conductivity, and the linear-in-frequency increase of
$\sigma_{1}^{\rm{inter}}(\nu)$ is restored with a larger slope [the
grey arrow in panel (b)]. At $\nu \geq 6000$ cm$^{-1}$ (0.8 eV), the
optical conductivity flattens out, forming a broad flat maximum. We
attribute it to the transitions between the almost parallel bands
along the $M$--$R$ line, shown as the dotted green arrows. The
maximum is not sharp because there are many other transitions with
comparable frequencies; see, e.g., the dotted green arrow between
the $\Gamma$ and R points. After the relatively flat region,
$\sigma_{1}^{\rm{inter}}(\nu)$ continues to rise [see
Fig.~\ref{overview}(b)], as more and more bands get involved in
optical transitions.

In Fig.~\ref{bands}(b), we also compare our results with the
effective-Hamiltonian calculations~\cite{Grushin2019} for the
contributions near the $\Gamma$ point. An extrapolation of these
calculations, originally performed for $\nu < 320$ cm$^{-1}$ (40
meV), to higher frequencies is shown as a solid purple line. The
experimental $\sigma_{1}^{\rm{inter}}(\nu)$ is generally steeper
than the results of these calculations. A very similar behavior of
the experimental conductivity versus such effective-Hamiltonian
calculations has also been reported in Ref.~\cite{Rees2019}.
Perhaps, at the lowest frequencies the match between the experiment
and the model is better, but our signal-to-noise ratio is not
sufficient for final conclusions (one should also remember that the
$\sigma_{1}^{\rm{inter}}(\nu)$ spectrum is obtained utilizing a
Drude-terms subtracting procedure). In any case, the mismatch can be
related to deviations of the bands from linearity even at low
energies~\cite{Chang2017, Tang2017, Li2019, Bradlyn}. This can be
clarified in more advanced band-structure-based optical-conductivity
calculations, which are beyond the scope of this
paper~\cite{Habe2019}.

Having established the connection between the features in the
experimental low-energy interband conductivity and the band
structure, we would like to add another note on the intraband
response. The exact shape of the free-carrier contribution is
obviously sample dependent, because, e.g., the impurity scattering
rate differ from sample to sample. However, if the total amount of
doping is low enough, the plasma frequency is fixed by the position
of the Fermi level, which, in turn, can be found within the band
structure calculation procedure. Such calculations~\cite{Li2019}
produced $\hbar\omega_{pl} = 1.344$~eV, in excellent agrement with
our result obtained above, $1.39 \pm 0.08$~eV. Furthermore, the
calculated and the observed spectral positions of the screened
plasma frequency (the zero-crossing points of $\varepsilon_{1}$)
also match each other very well, as can be seen in
Fig.~\ref{comparison}(b). This agreement of the plasma frequencies
also indicates a good quality of our sample in terms of low defect
and impurity concentration.

\begin{figure}
\centering
\includegraphics[width=3 cm]{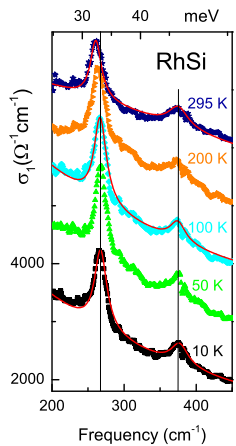}
\caption{Optical conductivity of RhSi at the frequencies near the
observed phonon modes for selected temperatures. The $y$ scale
corresponds to the 10-K spectra. The spectra for other temperatures
are shifted upwards for clarity. Results of Lorentzian fits of the
phonon modes (on an electronic background) are shown for 10, 100 and
295 K as solid red lines. The black vertical lines are a guide to
the eye.} \label{phonons}
\end{figure}

\subsection{Phonons}

Based on its crystallographic symmetry, RhSi is supposed to show
five infrared-active phonon modes; however, we can clearly identify
only two of them in our spectra. The other modes are likely too weak
to be resolved on top of the electronic background within the
available experimental accuracy. The two sharp phonon modes fall in
the spectral range from 200 to 400 cm$^{-1}$ (25 -- 50 meV); see
Fig.~\ref{phonons}(c). The positions of these modes at $T=10$~K are
marked with thin vertical lines.

Both modes can be accurately described by Lorentzians at any
temperature. No asymmetric (Fano-like) models are necessary. This is
in contrast to the situation in FeSi --~an isostructural analog of
RhSi with a presumably important role of electron correlations~--
where strong phonon-line asymmetry was reported in the optical
spectra and related to electron-phonon
coupling~\cite{Damascelli1997}. The absence of detectable
electron-phonon coupling indicates that phonon-mediated
electron-correlation effects are likely of no relevance in RhSi.

Let us finally mention that the phonon modes observed in RhSi
demonstrate a usual broadening as temperature rises. Additionally,
the low-frequency mode shows a small softening of its central
frequency with increasing $T$, which can be explained by usual
thermal expansion (possible softening of another mode might be not
resolved because of a larger width of this mode).

\section{Conclusions}

We have measured the broadband optical response of the multifold
semimetal RhSi. Infrared-active phonons and electronic transitions
are revealed in this study. The phonon modes demonstrate a trivial
temperature dependence with no indications of strong electron-phonon
coupling. The intraband electronic contribution (Drude) is
relatively strong with the (unscreened) plasma frequency of
approximately 1.4 eV. The interband optical conductivity
demonstrates a linear increase at low frequencies (below 300 meV).
We interpret this increase as a signature of the transitions between
the linear bands crossing around the $\Gamma$ point. At somewhat
higher frequencies (400 -- 600 meV), contributions from the linear
bands near the $R$ point (Pauli-blocked at lower frequencies)
manifest themselves as an increased slope of $\sigma_{1}(\omega)$.
These observations confirm the predictions for the
optical-conductivity behavior in multifold semimetals.

\section{Acknowledgements}

We thank Prof. Tetsuro Habe and Prof. Zhi Li for contacting us after
a preprint of this paper has been posted to the arxiv and for
subsequent useful discussions. We are grateful to Ms. Gabriele
Untereiner for valuable technical support. E.U. acknowledges the
support from the European Social Fund and from the Ministry of
Science, Research, and the Arts of Baden-W{\"u}rttemberg. This work
was partly supported by the Deutsche Forschungsgemeinschaft (DFG)
via Grant No. DR228/51-1.

\section{Supplemental material}

\begin{figure}
\centering
\includegraphics[width=\columnwidth,clip]{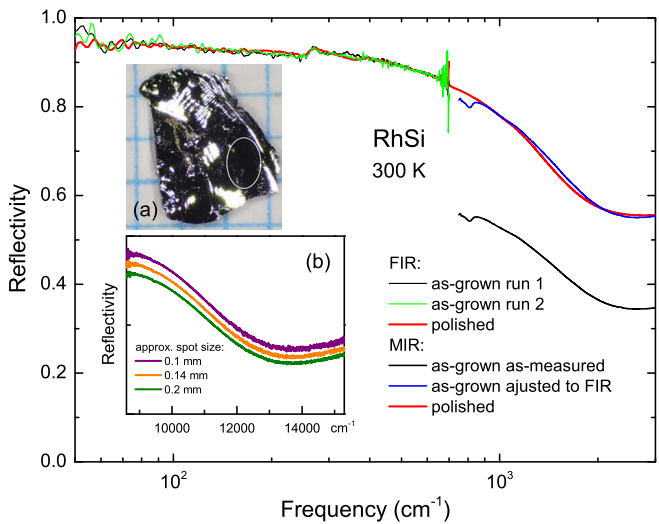}
\caption{Optical reflectivity of polished and as-grown RhSi samples
at 300 K [main frame], a photograph of the as-grown sample used in
these measurements [inset (a)], and reflectivity measured with
different apertures [inset (b)]. The far-infrared reflectivity (FIR,
up to 700 cm$^{-1}$) was measured utilizing an \textit{in situ} gold
coating technique, which allows measurements of samples with
irregular shapes~\cite{gold}. As one can see from the graph,
as-grown and polished surfaces provided identical results. At higher
frequencies -- in the mid-infrared (MIR) and above -- our equipment
does not enable gold evaporation. Thus, flat mirrors are used as
references. Due to the irregular sample shape, the absolute values
of reflectivity are lowered down as compared to the FIR measurements
(the wavy surface defocuses the reflected light). However, the shape
of the reflectivity curve (black line) is similar to the one
obtained on a polished sample (red line). By multiplying the
measured ``as-grown'' reflectivity with a constant coefficient, we
can reproduce the measurements performed on a polished surface (cf.
the red and blue curves) and to merge the MIR and FIR spectra. The
fact that defocusing is indeed essential is demonstrated in the
inset (b), where measurements made in an infrared microscope with a
few different spot sizes are shown. The measurements were performed
within the most flat part of the sample [white ellipse in the inset
(a)]. Flat mirrors were used as references and the spot size for the
reference measurements was the same as the spot size for the sample
measurements. One can see that the measurements performed with
smaller spot sizes show larger reflectivity. This proves that
defocusing due to the wavy sample surface is the reason for the
lower reflectivity values obtained on as-grown sample surfaces. We
note that even with the smallest spot size we could not reach the
reflectivity level of polished samples: even the best looking
portion of the as-grown sample surface is not sufficiently flat.}
\label{suppl1}
\end{figure}

\textbf{Justification of using polished samples. }The available
samples had irregular shape and could not be cleaved easily. We
collected the optical spectra on both, polished and as-grown,
surfaces. Comparison of the results of these measurements, provided
in Fig.~\ref{suppl1} and its caption, justifies that polished
samples can be utilized for reflectivity measurements.

\begin{figure}
\centering \vspace{0.5 cm}
\includegraphics[width=0.8\columnwidth]{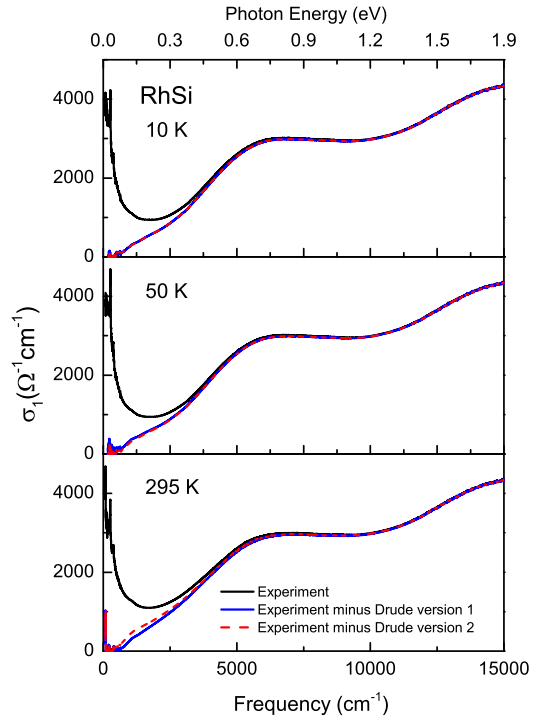}
\caption{As-measured optical conductivity of RhSi (real part) and
two versions of the interband part of conductivity obtained via
subtracting the Drude and phonon contributions as described in the
main text.} \label{suppl2}
\end{figure}

\textbf{Drude-term subtraction. }The Drude-term subtraction
procedure has become quite common in the
community~\cite{Chaudhuri2017, Corasaniti2019}. This method can be
helpful in extracting the interband contribution, but should be used
with care. Thus, we performed two independent versions of the
Lorentz-Drude fits for each measurement temperature and then
subtracted the intraband and phonon contributions from the
experimental curves. The results for 10, 50, and 295 K are shown in
Fig.~\ref{suppl2}. One can see that the linear-in-frequency
interband conductivity is robust: it survives independently of the
exact fit function (at 295 K, there are visible deviations from
linearity, but this is not unexpected at high temperatures).
Afterwards, we additionally varied the parameters of the Drude modes
(for each of the fit versions) within the uncertainty set by the
experimental spectra. We found that the approximate linearly of the
interband conductivity for $\nu < 3000$~cm$^{-1}$ did not change
with such variations.

\begin{thebibliography}{99}


\bibitem{Manes2012} J. L. Ma\~{n}es, Phys. Rev. B \textbf{85}, 155118 (2012).

\bibitem{Bradlyn2016} B. Bradlyn, J. Cano, Z. Wang, M. G. Vergniory, C. Felser,
R. J. Cava, B. A. Bernevig, Science \textbf{353}, aaf5037 (2016).

\bibitem{Chang2018} G. Chang, B. J. Wieder, F. Schindler, D. S. Sanchez,
I. Belopolski, S.-M. Huang, B. Singh, D. Wu, T.-R. Chang, T.
Neupert, S.-Y. Xu, H. Lin, and M. Z. Hasan, Nat. Mater. \textbf{17},
978 (2018).

\bibitem{Chang2017} G. Chang, S.-Y. Xu, B. J. Wieder, D. S. Sanchez, S.-M. Huang,
I. Belopolski, T.-R. Chang, S. Zhang, A. Bansil, H. Lin, and M. Z.
Hasan, Phys. Rev. Lett. \textbf{119}, 206401 (2017).

\bibitem{Tang2017} P. Tang, Q. Zhou, and S.-C. Zhang,
Phys. Rev. Lett. \textbf{119}, 206402 (2017).

\bibitem{Sanchez2019} D. S. Sanchez, I. Belopolski, T. A. Cochran, X. Xu, J.-X. Yin,
G. Chang, W. Xie, K. Manna, V. S\"u{\ss}, C.-Y. Huang, N. Alidoust,
D. Multer, S. S. Zhang, N. Shumiya, X. Wang, G.-Q. Wang, T.-R.
Chang, C. Felser, S.-Y. Xu, S. Jia, H. Lin, and M. Z. Hasan, Nature
\textbf{567}, 500 (2019).

\bibitem{Rao2019} Z. Rao, H. Li, T. Zhang, S. Tian, C. Li, B. Fu, C. Tang,
L. Wang, Z. Li, W. Fan, J. Li, Y. Huang, Z. Liu, Y. Long, C. Fang,
H. Weng, Y. Shi, H. Lei, Y. Sun, T. Qian, and H. Ding, Nature
\textbf{567}, 496 (2019).

\bibitem{Schroter2019} N. B. M. Schr\"{o}ter, D. Pei,
M. G. Vergniory, Y. Sun, K. Manna, F. de Juan, J.. A. Krieger, V.
S\"u{\ss}, M. Schmidt, P. Dudin, B. Bradlyn, T. K. Kim, T. Schmitt,
C. Cacho, C. Felser, V. N. Strocov, and Y. Chen, Nat. Phys.
\textbf{15}, 759 (2019).

\bibitem{Takane2019} D. Takane, Z. Wang, S. Souma, K. Nakayama,
T. Nakamura, H. Oinuma, Y. Nakata, H. Iwasawa, C. Cacho, T. Kim, K.
Horiba, H. Kumigashira, T. Takahashi, Y. Ando, and T. Sato, Phys.
Rev. Lett. \textbf{122}, 076402 (2019).

\bibitem{deJuan2017} F. de Juan, A. G. Grushin, T. Morimoto, and J. E. Moore,
Nat. Commun. \textbf{8}, 15995 (2017).

\bibitem{Rees2019} D. Rees, K. Manna, B. Lu, T. Morimoto, H. Borrmann, C. Felser,
J. E. Moore, D. H. Torchinsky, and J. Orenstein, arXiv:1902.03230.

\bibitem{Hosur2012} P. Hosur, S. A. Parameswaran, and A. Vishwanath,
Phys. Rev. Lett. \textbf{108}, 046602 (2012).

\bibitem{Bacsi2013} {\'{A}}. B{\'{a}}csi and A. Virosztek,
Phys. Rev. B \textbf{87}, 125425 (2013).

\bibitem{Ashby2014} P. E. C. Ashby and J. P. Carbotte,
Phys. Rev. B \textbf{89}, 245121 (2014).

\bibitem{Chen2015} R. Y. Chen, S. J. Zhang, J. A. Schneeloch,
C. Zhang, Q. Li, G. D. Gu, and N. L. Wang, Phys. Rev. B \textbf{92},
075107 (2015).

\bibitem{Xu2016} B. Xu, Y. M. Dai, L. X. Zhao, K. Wang, R. Yang,
W. Zhang, J. Y. Liu, H. Xiao, G. F. Chen, A. J. Taylor, D. A.
Yarotski, R. P. Prasankumar, and X. G. Qiu, Phys. Rev. B
\textbf{93}, 121110 (2016).

\bibitem{Neubauer2016} D. Neubauer, J. P. Carbotte, A. A. Nateprov,
A. L\"{o}hle, M. Dressel, and A. V. Pronin, Phys. Rev. B
\textbf{93}, 121202 (2016).

\bibitem{Kimura2017} S. Kimura, H. Yokoyama, H. Watanabe, J. Sichelschmidt,
V. S\"u{\ss}, M. Schmidt, and C. Felser, Phys. Rev. B \textbf{96},
075119 (2017).

\bibitem{Grushin2019} M.-\'{A}. S\'{a}nchez-Mart\'{i}nez, F. de Juan, and
A. G. Grushin, Phys. Rev. B \textbf{99}, 155145 (2019).

\bibitem{Li2019} Z. Li, T. Iitaka, H. Zeng, and H. Su, Phys. Rev. B \textbf{100},
155201 (2019).

\bibitem{gold} C. C. Homes, M. Reedyk, D. A. Cradles, and T. Timusk,
Appl. Opt. \textbf{32}, 2976 (1993).

\bibitem{Tanner2015} D. B. Tanner, Phys. Rev. B \textbf{91}, 035123 (2015).

\bibitem{Dressel2002} M. Dressel and G. Gr\"{u}ner,
{\it Electrodynamics of Solids} (Cambridge University Press,
Cambridge, 2002).

\bibitem{Schilling2017} M. B. Schilling, A. L\"{o}hle, D. Neubauer,
C. Shekhar, C. Felser, M. Dressel, and A. V. Pronin, Phys. Rev. B
\textbf{95}, 155201 (2017).

\bibitem{Neubauer2018} D. Neubauer, A. Yaresko, W. Li, A. L\"{o}hle,
R. H\"{u}bner, M. B. Schilling, C. Shekhar, C. Felser, M. Dressel,
and A. V. Pronin, Phys. Rev. B \textbf{98}, 195203 (2018).

\bibitem{Bradlyn} https://www.topologicalquantumchemistry.com; http://www.cryst.ehu.es;
B. Bradlyn, L. Elcoro, J. Cano, M. G. Vergniory, Z. Wang, C. Felser,
M. I. Aroyo, and B. A. Bernevig, Nature \textbf{547}, 298 (2017); M.
G. Vergniory, L. Elcoro, C. Felser, N. Regnault, B. A. Bernevig, and
Z. Wang, Nature \textbf{566}, 480 (2019).

\bibitem{Qiu2019} Z. Qiu, C. Le, Z. Liao, B. Xu, R. Yang, J. Hu, Y. Dai, and X. Qiu,
Phys. Rev. B \textbf{100}, 125136 (2019).

\bibitem{Hutt2018} F. H\"{u}tt, A. Yaresko, M. B. Schilling, C. Shekhar, C. Felser,
M. Dressel, and A. V. Pronin, Phys. Rev. Lett. \textbf{121}, 176601
(2018).

\bibitem{Kemmler2018} R. Kemmler, R. H\"{u}bner, A. L\"{o}hle, D. Neubauer,
I. Voloshenko, L. M. Schoop, M. Dressel and A. V. Pronin, J. Phys.:
Condens. Matter \textbf{30}, 485403 (2018).

\bibitem{Frenzel2017} A. J. Frenzel, C. C. Homes, Q. D. Gibson, Y. M. Shao,
K. W. Post, A. Charnukha, R. J. Cava, and D. N. Basov, Phys. Rev. B
\textbf{95}, 245140 (2017).

\bibitem{Chaudhuri2017} D. Chaudhuri, B. Cheng, A. Yaresko, Q. D. Gibson,
R. J. Cava, and N. P. Armitage, Phys. Rev. B \textbf{96}, 075151
(2017).

\bibitem{Grassano2018} D. Grassano, O. Pulci, A. M. Conte, F. Bechstedt,
Sci. Rep. \textbf{8}, 3534 (2018).

\bibitem{Habe2019} An elaborative theoretical study of the optical conductivity of CoSi
-- another compound from the same multifold family -- has been
recently published: T. Habe, Phys. Rev. B \textbf{100}, 245131
(2019).

\bibitem{Damascelli1997} A. Damascelli, K. Schulte, D. van der Marel, and A. A. Menovsky,
Phys. Rev. B \textbf{55}, R4863 (1997).

\bibitem{Corasaniti2019} M. Corasaniti, R. Yang, A. Pal, M. Chinotti, L. Degiorgi, A. Wang,
and C. Petrovic, Phys. Rev. B \textbf{100} 041107 (2019).

\end{thebibliography}
\end{document}